\documentclass[conference]{IEEEtran}
\usepackage[latin9]{inputenc}
\usepackage{amsmath}
\usepackage{amsthm}
\usepackage{graphicx}
\usepackage{amssymb}
\usepackage{subfig}
\usepackage{float}
\usepackage{cite}
\usepackage{caption}
\newtheorem{theorem}{Theorem}

\newtheorem{lemma}{Lemma}
\usepackage[ruled,vlined]{algorithm2e}

\ifCLASSINFOpdf
\else
\fi

\usepackage{stfloats}

\begin{document}

\title{Privacy-Aware Location Sharing \\ with Deep Reinforcement Learning}

\author{\IEEEauthorblockN{Ecenaz Erdemir, Pier Luigi Dragotti and Deniz G{\"u}nd{\"u}z}
\IEEEauthorblockA{Imperial College London \\Department of Electrical and Electronic Engineering\\
Email: \{e.erdemir17, p.dragotti, d.gunduz\}@imperial.ac.uk}}


\maketitle



\begin{abstract}
Location-based services (LBSs) have become widely popular. Despite their utility, these services raise concerns for privacy since they require sharing location information with untrusted third parties. In this work,
we study privacy-utility trade-off in location sharing mechanisms. Existing approaches are mainly focused on privacy of sharing a single location or myopic location trace privacy; neither of them taking into account the temporal correlations between the past and current locations. Although these methods preserve the privacy for the current time, they may leak significant amount of information at the trace level as the adversary can exploit temporal correlations in a trace.
We propose an information theoretically optimal privacy preserving location release mechanism that takes temporal correlations into account. We measure the privacy leakage by the mutual information between the user's true and released location traces. To tackle the history-dependent mutual information minimization, we reformulate the problem as a Markov decision process (MDP), and solve it using asynchronous actor-critic deep reinforcement learning (RL).

\end{abstract}


%
\IEEEpeerreviewmaketitle

\vspace{-0.2cm}
\section{Introduction}
Fast advances in mobile devices and positioning technologies have fostered the development of many location-based services (LBSs), such as Google Maps, Uber, Forsquare and Tripadvisor. These services provide users with useful information about their surroundings, transportation services, friends' activities, or nearby attraction points. Moreover, the integration of LBSs with social networks, such as Facebook, Twitter, YouTube, has rapidly increased indirect location sharing, e.g., via image or video sharing. However, location is one of the most sensitive private information for users, since a malicious adversary can use this information to derive users' habits, health condition, social relationships, or religion. Therefore, location trace privacy has been an important concern in LBSs, and there is an increasing pressure from consumers to keep their traces private against malicious attackers or untrusted service providers (SPs), while preserving the utility obtained from these applications.
\makeatletter{\renewcommand*{\@makefnmark}{}\footnotetext{This work was partially supported by the European Research Council (ERC) through project BEACON (No. 677854).}\makeatother}

A large body of research has focused on location-privacy protection mechanisms (LPPMs) against an untrusted service provider \cite{LPsurvey}. These methods can be categorized as spatial-location and temporal-location privacy preserving methods \cite{Classification}. While the former focuses on protecting a single location data \cite{Crypto,Shokri_kanony,Shokri_single,InfoTheo_single,Ravi}, the latter aims at providing location trace privacy \cite{Trace1,Trace2,Trace3}.
Individual locations on a trace are highly correlated, and the strategies focusing on the current location privacy might reveal sensitive information about the past or future locations. 

Differential privacy, k-anonymity and information theoretic metrics are commonly used as privacy measures \cite{Crypto,Shokri_kanony,Shokri_single,InfoTheo_single,Ravi,Trace1,Trace2,Trace3}.
By definition, differential privacy prevents the service provider from inferring the current location of the user, even if the SP has the knowledge of all the remaining locations. K-anonymity ensures that a location is indistinguishable from at least $k-1$ other location points. However, differential privacy and k-anonymity are meant to ensure the privacy of a single location, and they are shown not to be appropriate measures for location privacy in \cite{ShokriQuantify}. Instead, we treat the true and released location traces as random sequences, and measure the privacy leakage by mutual information \cite{PrivacyMetric}.

In \cite{Ravi}, the authors introduce location distortion mechanisms to keep the user's trajectory private. Privacy is measured by mutual information between the true and released traces and constrained by the average distortion for a specific distortion measure. The true trajectory is assumed to form a Markov chain. Due to the computational complexity of history-dependent mutual information optimization, authors propose bounds which take only the current and one step past locations into account. However, due to temporal correlations in the trajectory, the optimal distortion introduced at each time instance depends on the entire distortion and location history. Hence, the proposed bounds do not guarantee optimality.



In this work, we consider the scenario in which the user follows a trajectory generated by a first-order Markov process, and periodically reports a distorted version of her location to an untrusted service provider. We assume that the true locations become available to the user in an online manner. We use the mutual information between the true and distorted location traces as a measure of privacy loss. For the privacy-utility trade-off, we introduce an online LPPM minimizing the mutual information while keeping the distortion below a certain threshold. Unlike \cite{Ravi}, we consider location release policies which take the entire released location history into account, and show its optimality.
To tackle the complexity, we exploit the Markovity of the true user trajectory, and recast the problem as a Markov decision process (MDP). After identifying the structure of the optimal policy, we use advantage actor-critic (A2C) deep reinforcement learning (RL) framework as a tool to evaluate our continuous state and action space MDP numerically.


\vspace{-0.3cm}
\section{Problem Statement}

\vspace{-0.3cm}
We consider a user who shares her location with a service provider to gain utility through some LBS. We denote the true location of the user at time $t$ by $X_t \in \mathcal{W}$, where $\mathcal{W}$ is the finite set of possible locations. We assume that the user trajectory $\{X_t\}_{t\geq 1}$ follows a first-order time-homogeneous Markov chain with transition probabilities $q_x(x_{t+1}|x_t)$, and initial probability distribution $p_{x_1}$. At time $t$, the user shares a distorted version of her current location, denoted by $Y_t \in \mathcal{W}$, with the untrusted service provider due to privacy concerns. We assume that the user shares the distorted location in an online manner; that is, the released location at time $t$ does not depend on the future true locations; i.e., for any $1 < t< n$, $Y_t \rightarrow (X^t, Y^{t-1}) \rightarrow (X_{t+1}^n, Y_{t+1}^n)$ form a Markov chain, where we have denoted the sequence $(X_{t+1}, \dots, X_n)$ by $X_{t+1}^n$, and the sequence $(X_1, \dots, X_t)$ by $X^t$.

Our goal is to characterize the trade-off between the privacy and utility. We quantify privacy by the information leaked to the untrusted service provider, measured by the mutual information between the true and released location trajectories. The information leakage of the user's location release strategy for a time period $n$ is given by
\begin{align}
\hspace{-0.1cm}    I(X^n;Y^n) = \hspace{-0.1cm} \sum\limits_{t=1}^{n}I(X^n;Y_t|Y^{t-1}) = \hspace{-0.1cm}  \sum\limits_{t=1}^{n}I(X^t;Y_t|Y^{t-1}), \label{eq:2}
\end{align}
where the first equality follows from the chain rule of mutual information, and the second from the Markov chain $Y^{t}\rightarrow (X_t,Y^{t-1}) \rightarrow X_{t+1}^n$. 

Releasing distorted locations also reduces the utility received from the service provider. Therefore, the distortion applied by the user should be limited. The distortion between the true location $X_t$ and the released location $Y_t$ is measured by a specified distortion measure $d(X_t,Y_t)$ (e.g., Manhattan distance or Euclidean distance).

Our goal is to minimize the information leakage rate to the service provider while keeping the average distortion below a specified level for utility. The infinite-horizon optimization problem can be written as:

\vspace{-0.6cm}
\begin{align}
    \min_{\{q_t(y_t|x^t,y^{t-1})\}_{t=1}^{\infty}} & \lim_{n \rightarrow \infty} \frac{1}{n}\sum_{t=1}^{n}I^{\boldsymbol{q}}(X^t;Y_t|Y^{t-1}) \label{eq:4} \\ 
    \mbox{such that } & \lim_{n \rightarrow \infty} \mathbb{E}\Bigg[\frac{1}{n} \sum_{t=1}^n d(X_t, Y_t) \Bigg]\leq \bar{D}, \label{eq:4b} 
\end{align}
where $\bar{D}$ is the specified average distortion constraint on the utility loss, $x_t$ and $y_t$ represent the realizations of $X_t$ and $Y_t$, $q_t(y_t|x^t,y^{t-1})$ is a conditional probability distribution which represents the user's randomized \textit{location release policy} at time $t$. The
expectation in (\ref{eq:4b}) is taken over the joint probabilities of $X_t$ and $Y_t$, where the randomness stems from both the Markov process generating the true trajectory, and the random release mechanism $q_t(y_t|x^t,y^{t-1})$.  The mutual information induced by policy $q_t(y_t|x^t,y^{t-1})$ is calculated using the joint probability distribution
\begin{align}
    P^{\boldsymbol{q}}(X^n=x^n,Y^n & =y^n) = p_{x_1}q_1(y_1|x_1) \nonumber \\
    &\times \prod_{t=2}^n \big [ q_x(x_t|x_{t-1})q_t(y_t|x^t,y^{t-1}) \big ], \label{eq:jointH}
\end{align}
where $\boldsymbol{q}=\{q_t(y_t|x^t,y^{t-1})\}_{t=1}^n$.
In the next section, we characterize the structure of the optimal location release policy, and using this structure recast the problem as an MDP, and finally evaluate the optimal trade-off numerically using deep RL.

\vspace{-0.2cm}
\section{Privacy-utility trade-off for \\online location sharing}

In this section, we analyze the optimal privacy-utility trade-off achievable by an LPPM under the notion of mutual information minimization with a distortion constraint. Moreover, we propose simplified location release policies that still preserve the optimality.

By the definition of mutual information, the objective in (\ref{eq:4}) depends on the entire history of $X$ and $Y$. Therefore, the user must follow a history-dependent location release policy $q_t^h(y_t|x^t,y^{t-1})$, where a feasible set $\mathcal{Q}_H$ satisfies $\sum_{y_t \in \mathcal{W}}q_t^h(y_t|x^t,y^{t-1})=1$. As a result of strong history dependence, computational complexity of the minimization problem increases exponentially with the increasing length of user trajectory. To tackle this problem, we introduce a class of simplified policies.

\subsection{Simplified Location Release Policies}

In this section we introduce a set of policies $\mathcal{Q}_S\subseteq \mathcal{Q}_H$ of the form $q_t^s(y_t|x_t,x_{t-1},y^{t-1})$, which samples the distorted location only by considering the last two true locations and the entire released location history. Hence, the joint distribution (\ref{eq:jointH}) induced by $\boldsymbol{q}_s \in \mathcal{Q}_S$, where $\boldsymbol{q}_s=\{q_t^s(y_t|x_t,x_{t-1},y^{t-1})\}_{t=1}^n$ can be written as
\begin{align}
    P^{\boldsymbol{q}_s}(X^n= & x^n ,Y^n =y^n) = p_{x_1}q_1^s(y_1|x_1) \nonumber \\
    &\times \prod_{t=2}^n \big [ q_x(x_t|x_{t-1})q_t^s(y_t|x_t,x_{t-1},y^{t-1}) \big ].
\end{align}

Next, we show that considering location release policies in set $\mathcal{Q}_S$ is without loss of optimality.

\begin{theorem}
In the minimization problem (\ref{eq:4}), there is no loss of optimality in restricting the location release policies to the set of policies $\boldsymbol{q}_s \in \mathcal{Q}_S$. Furthermore, information leakage induced by any $\boldsymbol{q}_s \in \mathcal{Q}_S$ can be written as: 
\begin{align}
    &I^{\boldsymbol{q}_s}(X^n,Y^n) = \sum\limits_{t=1}^{n}I^{\boldsymbol{q}_s}(X_t,X_{t-1};Y_t|Y^{t-1})  \\
    &=\sum\limits_{t=1}^{n} \hspace{-0.3cm} \sum \limits_{\substack{y^t \in \mathcal{W}^t \\ (x_t,x_{t-1}) \in \mathcal{W}}}  \hspace{-0.6cm}
    P^{\boldsymbol{q}_s}(x_t,x_{t-1},y^t)\log \frac{q_t^s(y_t|x_t,x_{t-1},y^{t-1})}{P^{\boldsymbol{q}_s}(y_t|y^{t-1})}.
    \label{eq:theo1}
\end{align}
\label{prop:1}
\end{theorem}

\vspace{-0.2cm}
The proof of Theorem \ref{prop:1} relies on the following lemmas and will be presented later.

\begin{lemma}
For any $\boldsymbol{q} \in \mathcal{Q}_H$, 
\begin{align}
    I^{\boldsymbol{q}}(X^n;Y^n) \geq \sum \limits_{t=1}^{n} I^{\boldsymbol{q}}(X_t,X_{t-1};Y_t|Y^{t-1})
\end{align}
with equality if and only if $\boldsymbol{q} \in \mathcal{Q}_S$.
\label{lem:1}
\end{lemma}

\begin{IEEEproof}
For any $\boldsymbol{q} \in \mathcal{Q}_H$,
    \begin{align}
        I^{\boldsymbol{q}}(X^n;Y^n) &= \sum\limits_{t=1}^{n}I^{\boldsymbol{q}}(X^t;Y_t|Y^{t-1}) \label{eq:10}\\
        &\geq \sum\limits_{t=1}^{n}I^{\boldsymbol{q}}(X_t,X_{t-1};Y_t|Y^{t-1}), \label{eq:11}
    \end{align}
where (\ref{eq:10}) follows from (\ref{eq:2}), and (\ref{eq:11}) from the non-negativity of mutual information. 
\end{IEEEproof}

\begin{lemma}
For any $\boldsymbol{q}_h \in \mathcal{Q}_H$, there exists a $\boldsymbol{q}_s \in \mathcal{Q}_S$ such that
\begin{align}
    \sum \limits_{t=1}^{n} I^{\boldsymbol{q}_h}(X_t,X_{t-1};Y_t|Y^{t-1}) = \sum \limits_{t=1}^{n} I^{\boldsymbol{q}_s}(X_t,X_{t-1};Y_t|Y^{t-1}).
\end{align}
\label{lem:2}
\end{lemma}

\vspace{-0.4cm}
\begin{IEEEproof}
For any $\boldsymbol{q}_h \in \mathcal{Q}_H$,  we choose the policy $\boldsymbol{q}_s \in \mathcal{Q}_S$ such that 
    \begin{align}
         \hspace{-0.3cm} q_t^s(y_t|x_t,x_{t-1},y^{t-1}) \hspace{-0.1cm} =  \hspace{-0.1cm} P^{\boldsymbol{q}_h}_{Y_t|X_t,X_{t-1},Y^{t-1}}(y_t|x_t,x_{t-1},y^{t-1}),  \hspace{-0.2cm} \label{eq:13}
    \end{align}
and we show that $P^{\boldsymbol{q}_h}_{X_t,X_{t-1},Y^{t}}=P^{\boldsymbol{q}_s}_{X_t,X_{t-1},Y^{t}}$. Then, $I^{\boldsymbol{q}_h}(X_t,X_{t-1};Y_t|Y^{t-1})=I^{\boldsymbol{q}_s}(X_t,X_{t-1};Y_t|Y^{t-1})$ holds, which proves the statement in Lemma \ref{lem:2}.
The proof of $P^{\boldsymbol{q}_h}_{X_t,X_{t-1},Y^{t}}$=$P^{\boldsymbol{q}_s}_{X_t,X_{t-1},Y^{t}}$ is derived by induction as follows,
    \begin{align}
        & P^{\boldsymbol{q}_h}(x_{t+1},x_t,y^t) \nonumber \\ 
        & = \hspace{-0.3cm} \sum_{ \substack{ x_{t-1} \in \mathcal{W}}} \hspace{-0.3cm}  q_x(x_{t+1}|x_t)q_t^h(y_t|x_t,x_{t-1},y^{t-1}) P^{\boldsymbol{q}_h}(x_{t},x_{t-1},y^{t-1}) \nonumber \\
        & = \hspace{-0.3cm} \sum_{ \substack{ x_{t-1} \in \mathcal{W} }} \hspace{-0.3cm}  q_x(x_{t+1}|x_t)q_t^s(y_t|x_t,x_{t-1},y^{t-1})P^{\boldsymbol{q}_s}(x_{t},x_{t-1},y^{t-1}) \nonumber\\
        & = P^{\boldsymbol{q}_s}(x_{t+1},x_t,y^t),
    \end{align}
where (\ref{eq:13}) holds, and $P^{\boldsymbol{q}_h}_{X_1}(x)=p_{x_1}(x)=P^{\boldsymbol{q}_s}_{X_1}(x)$ is for the initialization of the induction.    
\end{IEEEproof}
\vspace{0.5cm}
\begin{IEEEproof}[Proof of Theorem \ref{prop:1}]
Following Lemmas \ref{lem:1} and \ref{lem:2}, for any $\boldsymbol{q}_h \in \mathcal{Q}_H$, there exists a $\boldsymbol{q}_s \in \mathcal{Q}_S$ such that \begin{align}
    I^{\boldsymbol{q}_h}(X^n;Y^n) \geq I^{\boldsymbol{q}_s}(X^n;Y^n).
\end{align}

Hence, there is no loss of optimality in using the location release policies of the form $q_t^s(y_t,|x_t,x_{t-1},y^{t-1})$, and information leakage reduces to (\ref{eq:theo1}).
\end{IEEEproof}

\begin{figure}[pt]
\centering
\includegraphics[width=8.7cm,height=2.5cm]{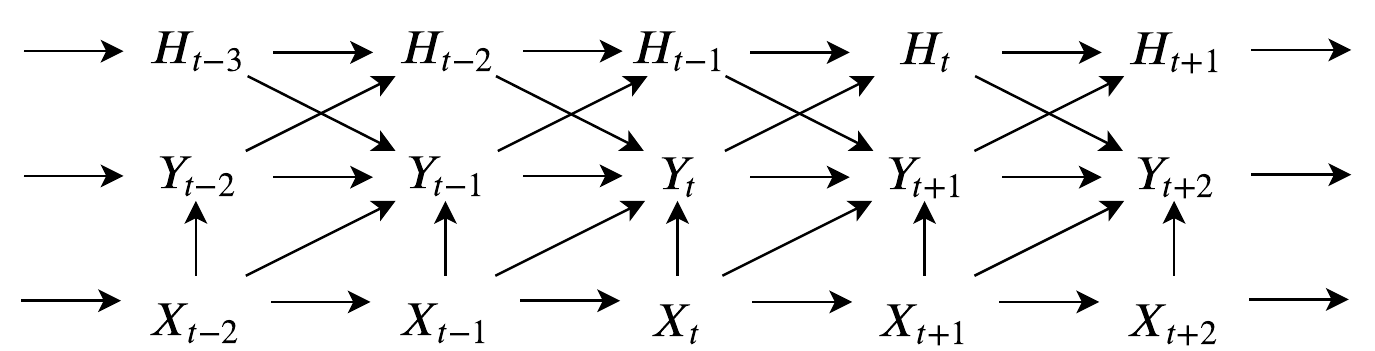}
\caption{Markov chain for the simplified location release policy.} 
\label{fig:MC}
\end{figure}

Restricting our attention to the user location release policies $\boldsymbol{q}_s \in \mathcal{Q}_S$, we can write the minimization problem (\ref{eq:4}) as
\begin{align}
    \min_{\boldsymbol{q}_s:\{\mathbb{E}^{\boldsymbol{q}_s}[d(x_t,y_t)]\leq \bar{D}\}_{i=1}^{n}} \hspace{0.1cm} \frac{1}{n} \sum_{t=1}^{n}I^{\boldsymbol{q}_s}(X_t,X_{t-1};Y_t|Y^{t-1}). \label{eq:Objective}
\end{align}

The location release strategy followed by the user is illustrated by the Markov chain in Fig. \ref{fig:MC}, where $H_t$ denotes the released location history, i.e., $H_t=Y^t$. That is, the user samples a distorted location, $Y_t$, at time t by considering the current and previous true locations, $(X_t,X_{t-1})$, and released location history, $(H_{t-2},Y_{t-1})$. 

Minimization of the mutual information subject to a utility constraint can be converted into an unconstrained minimization problem using Lagrange multipliers. Since the distortion constraint is memoryless, we can integrate it into the additive objective function easily. Hence, the unconstrained minimization problem for online location release privacy-utility trade-off can be rewritten as

\vspace{-0.4cm}
\begin{align}
    \min_{\boldsymbol{q}_s \in \boldsymbol{Q}_s} \hspace{-0.1cm} \frac{1}{n} \sum_{t=1}^{n} \hspace{-0.1cm}  I^{\boldsymbol{q}_s}(X_t,X_{t-1};Y_t|Y^{t-1}) + \lambda (\mathbb{E}^{\boldsymbol{q}_s}[d(x_t,y_t)]- \bar{D}). \label{eq:UnconsObjective}
\end{align}

\vspace{-0.6cm}
\subsection{MDP Formulation}
Markovity of the user's true location trace and the additive objective function in (\ref{eq:UnconsObjective}) allow us to represent the problem as an MDP with state $X_t$. However, the information leakage at time $t$ depends on $Y^{t-1}$, resulting in a growing state space in time. Therefore, for a given policy $\boldsymbol{q}_s$ and any realization $y^{t-1}$ of $Y^{t-1}$, we define a belief state $\beta_t \in \mathcal{P}_X$ as a probability distribution over the state space:
\begin{align}
    \beta_t(x_{t-1})=P^{\boldsymbol{q}_s}(X_{t-1}=x_{t-1}|Y^{t-1}=y^{t-1}).
\end{align}

This represents the service provider's belief on the user's true location at the beginning of time instance $t$, i.e., after receiving the distorted location $y_{t-1}$ at the end of the previous time  instance $t-1$. The MDP actions are defined as the probability distributions sampling the released location $Y_t=y_t$ at time $t$, and determined by the randomized location release policies. The user's action induced by a policy $\boldsymbol{q}_s$ can be denoted by $a_t(y_t|x_t,x_{t-1})=P^{\boldsymbol{q}_s}(Y_t=y_t|X_t=x_t,X_{t-1},\beta_t)$ \cite{AshishInfoTheoP,Giulio,ICASSP}. At each time $t$, the service provider updates its belief on the true location $\beta_{t+1}(x_t)$ after observing the distorted location $y_t$ by

\vspace{-0.5cm}
\begin{align}
    \beta&_{t+1}(x_{t}) =\frac{p(x_{t},y_t|y^{t-1})}{p(y_t|y^{t-1})}= \frac{\sum_{x_{t-1}}p(x_{t},x_{t-1},y_t|y^{t-1})}{\sum_{x_t,x_{t-1}}p(x_{t},x_{t-1},y_t|y^{t-1})} \nonumber \\
    & = \frac{\sum_{x_{t-1}}p(x_{t}|x_{t-1})q_t^s(y_t|x_t,x_{t-1},y^{t-1})p(x_{t-1}|y^{t-1})}{\sum_{x_t,x_{t-1}}p(x_{t}|x_{t-1})q_t^s(y_t|x_t,x_{t-1},y^{t-1})p(x_{t-1}|y^{t-1})} \nonumber\\
    &= \frac{\sum_{x_{t-1}}q_x(x_{t}|x_{t-1})a(y_t|x_t,x_{t-1})\beta_t(x_{t-1})}{\sum_{x_t,x_{t-1}}q_x(x_{t}|x_{t-1})a(y_t|x_t,x_{t-1})\beta_t(x_{t-1})}.
    \label{eq:BeliefUpdate}
\end{align}
We define per-step information leakage of the user due to taking action $a_t(y_t|x_t,x_{t-1})$ at time $t$ as,

\begin{align} 
\label{eq:PerStepObjective}
 l_t(x_t,x_{t-1},a_t,y^t;\boldsymbol{q}_s) := \log \frac{a_t(y_t|x_t,x_{t-1})}{P^{\boldsymbol{q}_s}(y_t|y^{t-1})}.
\end{align}

\sloppy
The expectation of $n$ step sum of (\ref{eq:PerStepObjective}) over the joint probability $P^{\boldsymbol{q}_s}(X_t,X_{t-1},Y^t)$ is equal to the mutual information expression in the original problem (\ref{eq:Objective}). Therefore, given belief and action probabilities, average information leakage at time $t$ can be formulated as,
\begin{align}
\mathbb{E}^{\boldsymbol{q}_s}[l_t(x^t_{t-1},&a_t,y^t)]  \hspace{-0.05cm} = \hspace{-0.8cm}
\sum_{ \substack{x_t,x_{t-1},y_t \in \mathcal{W}}}  \hspace{-0.7cm}   \beta_t(x_{t-1})a_t(y_t|x_t,x_{t-1})q_x(x_{t}|x_{t-1}) \nonumber \\  
& \times \log \frac{a_t(y_t|x_t,x_{t-1})}{\hspace{-0.5cm}\sum\limits_{\substack{\hat{x}_t,\hat{x}_{t-1} \in \mathcal{W}}} \hspace{-0.5cm}  \beta_t(\hat{x}_{t-1})a_t(y_t|\hat{x}_t,\hat{x}_{t-1})q_x(\hat{x}_{t}|\hat{x}_{t-1})}\nonumber \\
&:=\mathcal{L}(\beta_t,a_t).
\label{eq:AvgLeakage}
\end{align}

We remark that the representation of average distortion in terms of belief and action probabilities is straightforward due to its additive form. Similarly to (\ref{eq:AvgLeakage}), average distortion at time $t$ can be written as,
\begin{align}
    \mathbb{E}^{\boldsymbol{q}_s}[d(x_t,y_t)] \hspace{-0.05cm} &= \hspace{-0.8cm}  \sum_{x_t,x_{t-1},y_t\in \mathcal{W}} \hspace{-0.70cm}  \beta_t(x_{t-1} \hspace{-0.05cm} )a_t(y_t|x_t,x_{t-1} \hspace{-0.05cm} )q_x( \hspace{-0.05cm} x_t|x_{t-1} \hspace{-0.05cm} )d(x_t,y_t \hspace{-0.05cm} ) \nonumber \\
    &:= \mathcal{D}(\beta_t,a_t)
\end{align}

Finally, we can recast the original problem in (\ref{eq:UnconsObjective}) as a continuous state and action space MDP. Evaluation of the MDP relies on minimizing the objective 
\begin{align}
    \mathcal{C}(\beta_t,a_t)=\mathcal{L}(\beta_t,a_t)+\lambda (\mathcal{D}(\beta_t,a_t)-\bar{D}) \label{eq:MDPCost}
\end{align}
at each time step $t$ for a trajectory of length $n$.

Finding optimal policies for continuous state and action space MDPs is a PSPACE-hard problem \cite{PSPACEhard}.
In practice, they can be solved by various finite-state MDP evaluation methods, e.g., value iteration, policy iteration and gradient-based methods. These are based on the discretization of the continuous belief states to obtain a finite state MDP \cite{Tamas}. While finer discretization of the belief reduces the loss from the optimal solution, it causes an increase in the state space; hence, in the complexity of the problem. Therefore, we use a deep learning based method as a tool to numerically solve our continuous state and action space MDP problem.

\subsection{Advantage Actor-Critic (A2C) Deep RL}

In RL, an agent discovers the best action to take in a particular state by receiving instantaneous rewards/costs from the environment \cite{SuttonBarto}. On the other hand, in our problem, we have the knowledge of state transitions and the cost for every state-action pair without a need for interacting with the environment. We use A2C-deep RL as a computational tool to numerically evaluate the optimal location release policies for our continuous state and action space MDP. 

\begin{figure}[pt]
\centering
\includegraphics[width=6.6cm]{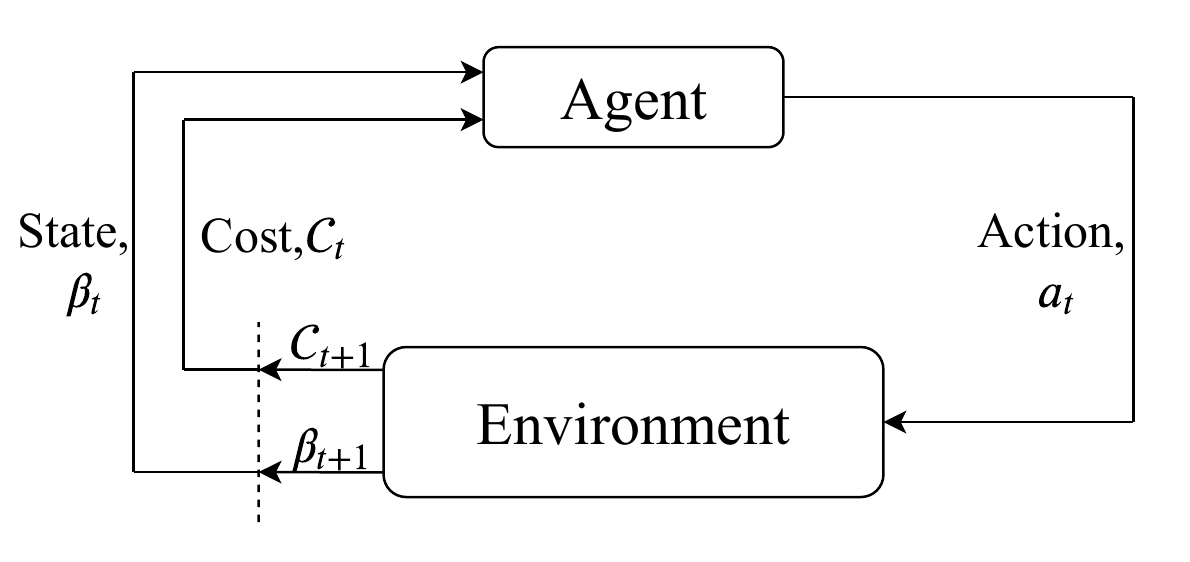}
\caption{RL for a known model.} 
\label{fig:RL}
\end{figure}

To integrate RL framework into our problem, we create an artificial environment which inputs the user's current action, $a_t(y_t|x_t,x_{t-1})$, samples an observation $y_t$, and calculates the next state, $\beta_{t+1}$, using Bayesian belief update (\ref{eq:BeliefUpdate}). Instantaneous cost revealed by the environment is calculated by (\ref{eq:MDPCost}). The user receives the experience tuple $(\beta_t,a_t,y_t,\beta_{t+1},\mathcal{C}_t)$ from the environment, and refines her policy accordingly.
Fig. \ref{fig:RL} illustrates the interaction between the artificial environment and the user, which is represented by the RL agent.
The corresponding Bellman equation induced by the location release policy ${\boldsymbol{q}_s}$ can be written as 

\vspace{-0.4cm}
\begin{align}
    V^{\boldsymbol{q}_s}(\beta)+J({\boldsymbol{q}_s})= \min_{a}\Big\{ \mathcal{C}(\beta,a)+V^{\boldsymbol{q}_s}(\beta') \Big\},
    \label{eq:Bellman}
\end{align}
where $V^{\boldsymbol{q}_s}(\beta)$ is the state-value function, $\beta'$ is the updated belief state according to (\ref{eq:BeliefUpdate}), $a$ represents action probability distributions, and $J({\boldsymbol{q}_s})$ is the cost-to-go function, i.e., the expected future cost induced by policy ${\boldsymbol{q}_s}$ \cite{Bertsekas}.

RL methods can be divided into three groups: value-based, policy-based, and actor-critic \cite{OnACalgs}. Actor-critic methods combine the advantages of value-based (critic-only) and policy-based (actor-only) methods, such as low variance and continuous action producing capability. The actor represents the policy structure, while the critic estimates the value function \cite{SuttonBarto}. In our setting, we parameterize the value function by the parameter vector $\theta \in \Theta$ as $V_{\theta}(\beta)$, and the stochastic policy by $\xi \in \Xi$ as $q_{\boldsymbol{\xi}}$. The difference between the right and the left hand side of (\ref{eq:Bellman}) is called temporal difference (TD) error, which represents the error between the critic's estimate and the target differing by one-step in time \cite{SurveyACRL}. The TD error for the experience tuple $(\beta_t,a_t,y_t,\beta_{t+1},\mathcal{C}_t)$ is estimated as

\vspace{-0.4cm}
\begin{align}
    \delta_t=\mathcal{C}_t(\beta_t)+\gamma V_{\theta_t}(\beta_{t+1})-V_{\theta_t}(\beta_t),
\end{align}
where $\mathcal{C}_t(\beta_t)+\gamma V_{\theta_t}(\beta_{t+1})$ is called the TD target, and $\gamma$ is a discount factor that we choose very close to $1$ to approximate the Bellman equation in (\ref{eq:Bellman}) for our infinite-horizon average cost MDP. To implement RL in the infinite-horizon problem, we take sample averages over independent finite trajectories, which are generated by experience tuples at each time $t$ via Monte-Carlo roll-outs.

Instead of using value functions in actor and critic updates, we use advantage function to reduce the variance in policy gradient methods. The advantage can be approximated by TD error. Hence, the critic is updated by gradient descent as:
\begin{align}
    \theta_{t+1}=\theta_t+\alpha_t^c \nabla_{\theta}\ell_{c}(\theta_t),
\end{align}
where $\ell_c(\theta_t)=\delta_t^2$ is the critic loss and $\alpha_t^c$ is the learning rate of the critic at time $t$. The actor is updated similarly as, 
\begin{align}
    \xi_{t+1}=\xi_t - \alpha_t^a \nabla_{\xi}\ell_a(\xi_t),
\end{align}
where $\ell_a(\xi_t)=-\ln(q_s(y_t|\beta_t,\xi_t))\delta_t$ is the actor loss and $\alpha_t^a$ is the actor's learning rate. This method is called \textit{advantage actor-critic RL}.

\begin{figure}[pt]
\centering
\hspace{-0.8cm}
\subfloat{(a)}{\includegraphics[width=7cm]{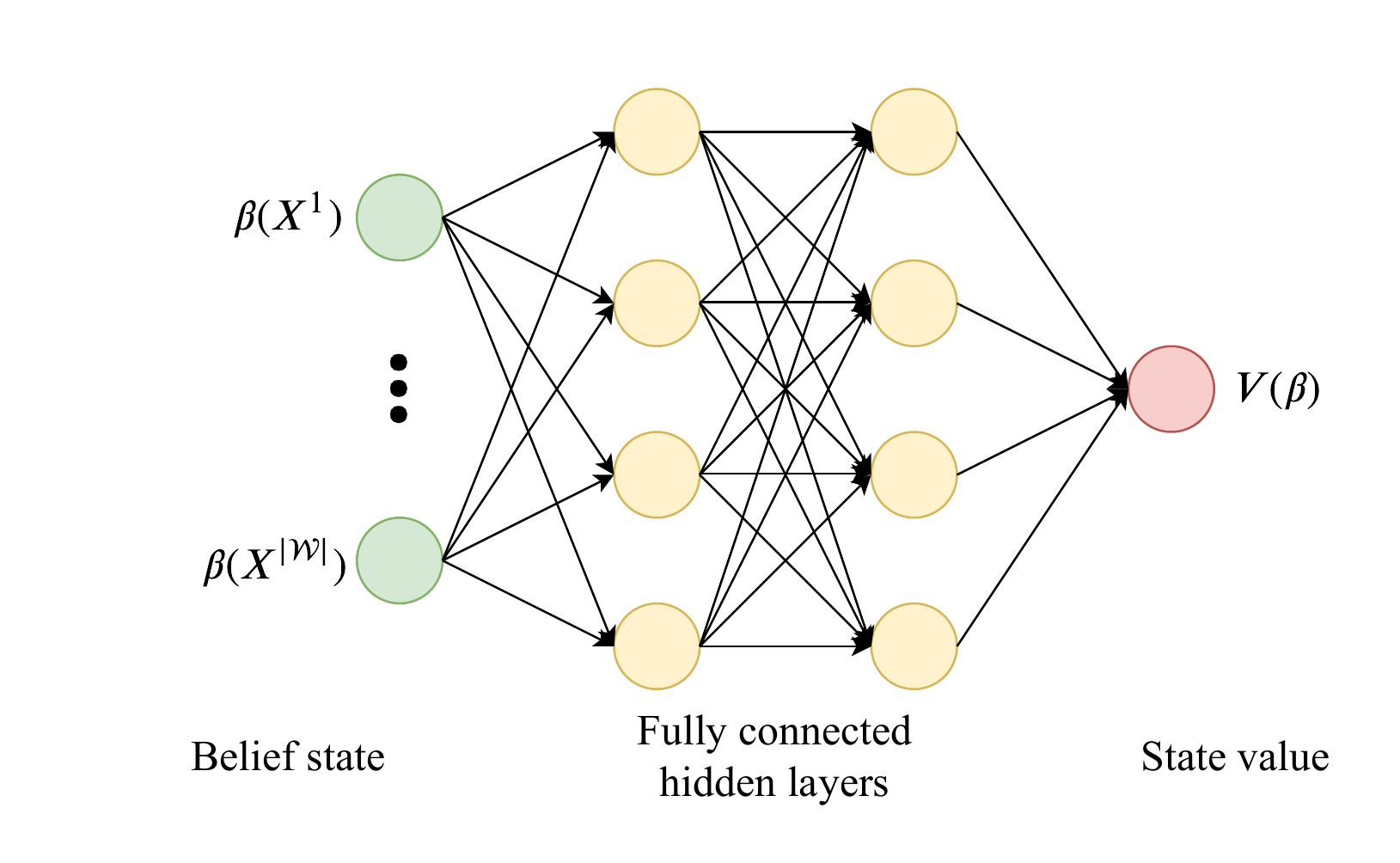}%
\label{fig:DeepCritic}}
\vfill
\subfloat{(b)}{\includegraphics[width=7.3cm]{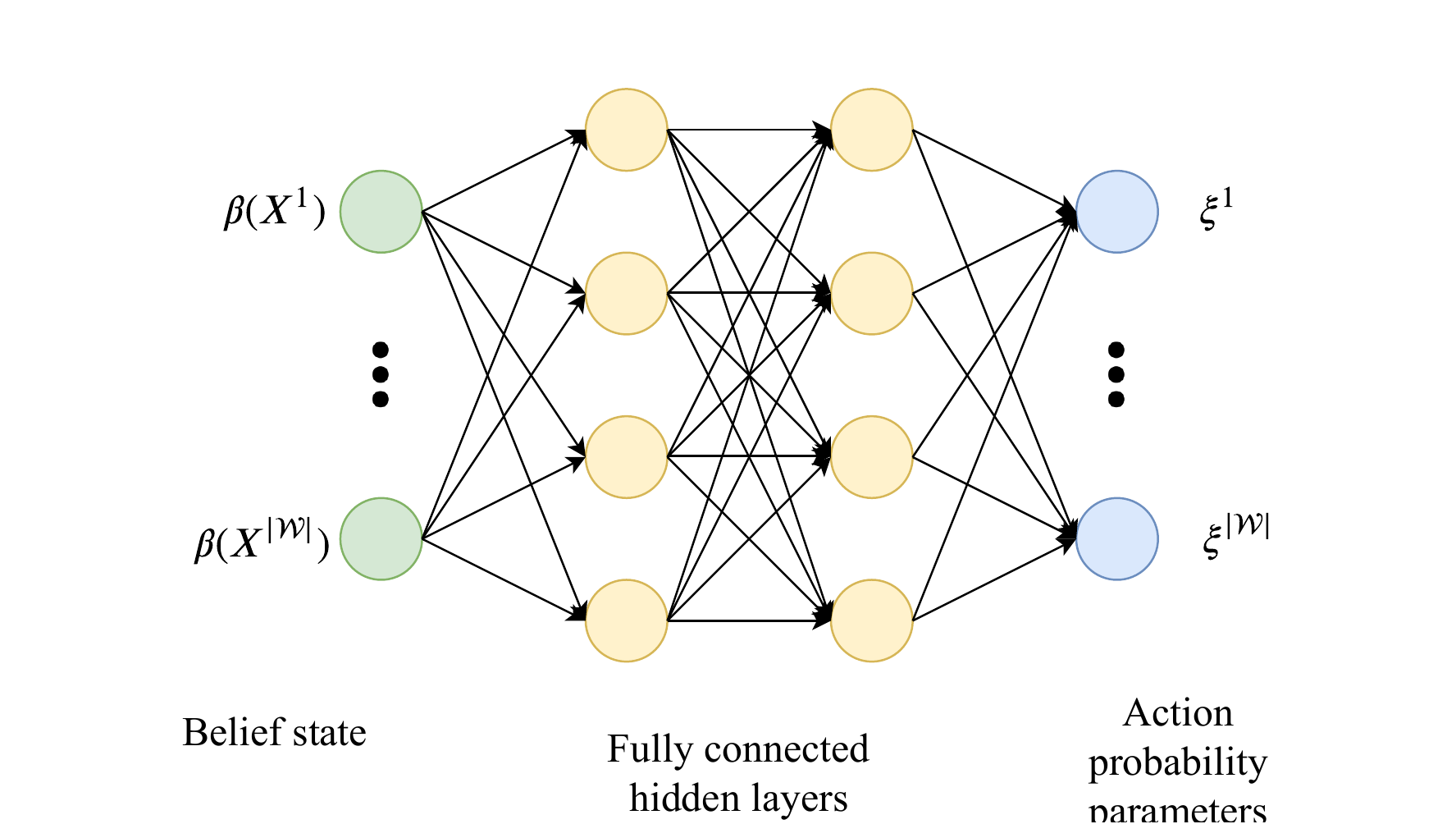}%
\label{fig:DeepActor}}
\caption{Critic (a) and actor (b) neural network structures.}
\label{fig:DeepNN}
\end{figure}

In our A2C-deep RL implementation, we represent the actor and critic mechanisms by fully connected feed-forward deep neural networks (DNNs) with two hidden layers as illustrated in Fig. \ref{fig:DeepNN}. The critic DNN takes the current belief state $\beta(\boldsymbol{X})$ as input, where $\boldsymbol{X}$ is the location vector of size $|\mathcal{W}|$, and outputs the value of the belief state for the current action probabilities $V_{\theta}^{\xi}(\beta)$. The actor takes the belief state as input, and outputs the parameters used for determining the action probabilities of the corresponding belief. Here, $\{\xi^1, \dots, \xi^{|\mathcal{W}|}\}$ are the concentration parameters of a Dirichlet distribution which represent the action probabilities. The overall A2C deep RL algorithm for online LPPM is described in Algorithm \ref{alg:A2CDRL}.

\begin{algorithm}[pt]
\SetAlgoLined
 Initialize the DNNs with random weights $\xi$ and $\theta$ \\
 Initialize environment $E$\\
 \For{episode=$1,N$}{
  Initialize belief state $\beta_0$\;
  \For{$t=0,n$}{
   Sample action probability vector $a_t \sim Dirichlet(a|\xi)$\ from the current policy;\\
   Perform the action and calculate cost $\mathcal{C}_{\xi_t}$ in $E$;\\
   Sample an observation $y_t$ and calculate the next belief state $\beta_{t+1}$ in $E$;\\
   Set TD target $\mathcal{C}_{\xi_t}+\gamma V_{\theta_t}^{\xi}(\beta_{t+1})$;\\
   Minimize the loss $\ell_c(\theta)=\delta^2=(\mathcal{C}_{\xi_t}+\gamma V_{\theta_t}^{\xi}(\beta_{t+1})-V_{\theta_t}^{\xi}(\beta_t))^2$;\\
   Update the critic $\theta \leftarrow \theta + \alpha^c \nabla_{\theta}\delta^2$;\\
   Minimize the loss $\ell_a(\xi_t)=(\ln(Dirichlet(a|\xi_t))\delta_t)$;\\
   Update the actor $\xi \leftarrow \xi -\alpha^a \nabla_{\xi}\ell_a(\xi_t)$;\\
   Update the belief state $\beta_{t+1} \leftarrow \beta_t$ 
   }{
  }
 }
 \caption{A2C-deep RL algorithm for online LPPM} \label{alg:A2CDRL}
\end{algorithm}

\vspace{-0.2cm}
\section{Numerical Results}

In this section, we evaluate the performance of the proposed LPPM policy for a simple grid-world example, and compare the results with the myopic Markovian location release mechanism proposed in \cite{Ravi}. In \cite{Ravi}, an upper bound on the privacy-utility trade-off is given by a myopic policy as follows:

\vspace{-0.5cm}
\begin{align}
    \sum \limits_{t=1}^{n} \min_{\substack{q(y_t|x_t,x_{t-1},y_{t-1}):\\\mathbb{E}^q[d(x_t,y_t)]\leq \bar{D}}} I^q(X_t,X_{t-1};Y_t|Y_{t-1}). \label{eq:RaviCost}
\end{align}
Exploiting the fact that (\ref{eq:RaviCost}) is similar to the rate-distortion function, Blahut-Arimoto algorithm is used in \cite{Ravi} to minimize the conditional mutual information at each time step. Finite-horizon solution of the objective function (\ref{eq:RaviCost}) is obtained by applying alternating minimization sequentially. In our simulations, we obtained the average information leakage and distortion for this approach by normalizing for $n=300$.

We consider a simple $4\times4$ grid-world, where $|\mathcal{W}|$=$16$. The cells are numbered such that the first and the last rows of the grid-world are represented by $\{1,2,3,4\}$ and $\{13,14,15,16\}$, respectively. User's trajectory forms a first-order Markov chain with the transition probability matrix $\boldsymbol{Q_x}$. The user can start its movement at any square with equal probability $p_{x_1}=\frac{1}{16}$. The Lagrangian multiplier $\lambda \in [0,20]$ denotes the user's choice of privacy-utility balance. 
We train two fully connected feed-forward DNNs, representing the actor and critic, by utilizing ADAM optimizer \cite{ADAM}. Both networks contain two hidden layers with leaky-ReLU activation \cite{LeakyRELU}. Distortion is measured by the Manhattan distance between $x_t$ and $y_t$. We obtain the corresponding privacy-utility trade-off by averaging the total information leakage and distortion over a horizon of $n=300$.

\sloppy
In Fig. \ref{fig:Sim}, privacy-distortion trade-off curves are obtained assuming that $\boldsymbol{Q}_x^0$, $\boldsymbol{Q}_x^1$ and $\boldsymbol{Q}_x^2$ are $16\times16$ Markov transition matrices with different correlation levels. In all three cases, the user can move from any square to any square at each step. While all the transition probabilities are equal, i.e. $\frac{1}{|\mathcal{W}|}$, for $\boldsymbol{Q}_x^0$, the probability of the user moving to a closer square is greater than taking a larger step to a more distant one for $\boldsymbol{Q}_x^1$ and $\boldsymbol{Q}_x^2$.
Moreover, $\boldsymbol{Q}_x^1$ represents a more uniform trajectory, where the agent moves to equidistant cells with equal probability, while with $\boldsymbol{Q}_x^2$ the agent is more likely to follow a certain path, i.e., the random trajectory generated by $\boldsymbol{Q}_x^2$ has lower entropy. The transition probabilities with $\boldsymbol{Q}_x^1$ are given by:

\begin{figure}[pt]
\centering
\includegraphics[width=8.85cm]{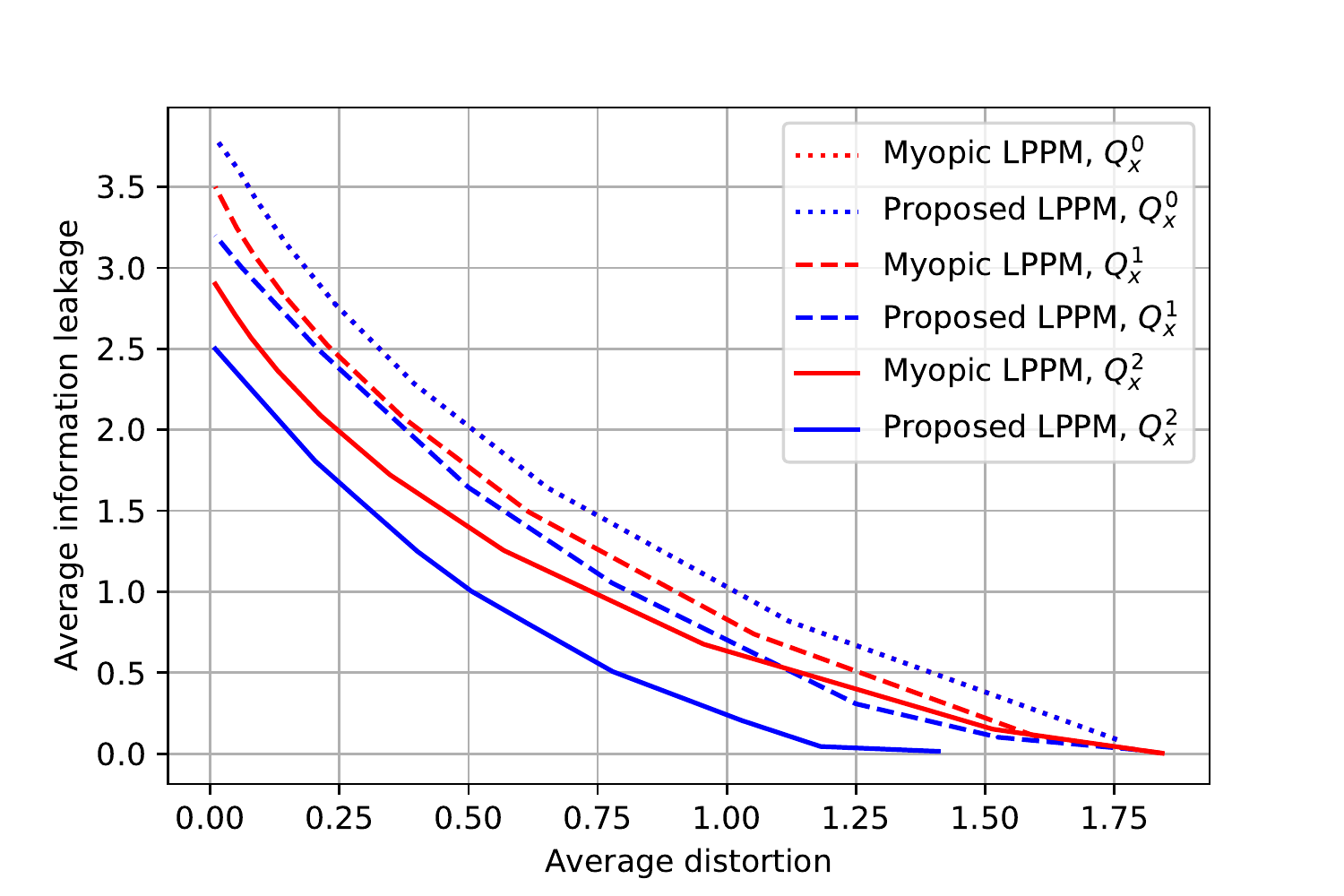}
\caption{Average information leakage as a function of the average distortion for the myopic and proposed LPPM policies.}
\label{fig:Sim}
\end{figure}

\vspace{-0.5cm}
\begin{align}
    q_x(x_t|x_{t+1})= \dfrac{ {r_{d(x_t,x_{t+1})}}/{d(x_t,x_{t+1})}  }{ \sum_{x_{t+1}}  {r_{d(x_t,x_{t+1})}}/{d(x_t,x_{t+1})}  } ,
\end{align}
where $d(x_t,x_{t+1})$ is the Manhattan distance between the user positions at time $t$ and $t+1$; $r_{d(x_t,x_{t+1})}$ is a scalar which determines the probability of the user moving from one grid to the equidistant grids in the next step. Fig. \ref{fig:Sim} is obtained by setting $r_0=1$ and $r_i=7-i$, $i=1,\dots,6$. Furthermore, we generate $\boldsymbol{Q}_x^2$ such that $q_x(x_t|x_{t+1})= \frac{ {u(x_t,x_{t+1})}/{d(x_t,x_{t+1})}  }{ \sum_{x_{t+1}}  {u(x_t,x_{t+1})}/{d(x_t,x_{t+1})}  } $, where, for $x_t \in \{1,2, \dots, 15\}$,

\vspace{-0.3cm}
\begin{align}
   u(x_t,x_{t+1})\hspace{-0.1cm} = \hspace{-0.1cm} \begin{cases}
    r_1, & \text{for $mod(x_t,4) \neq 0$, $x_{t+1}=x_t+1$}\\
    r_1, & \text{for $mod(x_t,4)=0$, $x_{t+1}=x_t+4$}, \\
    r_0, & \text{otherwise},
 \nonumber \end{cases}
\end{align}
$u(16,x_{t+1})=r_0$ for $x_{t+1} \in \{1, \dots, 15\}$, and $u(16,16)=r_1$. As a result, temporal correlations in the location history increase in the order $\boldsymbol{Q}_x^0$, $\boldsymbol{Q}_x^1$, $\boldsymbol{Q}_x^2$.

We train our DNNs for a time horizon of $n=300$ in each episode, and over $5000$ Monte Carlo roll-outs. Fig. \ref{fig:Sim} shows that, for $\boldsymbol{Q}_x^2$ the proposed LPPM obtained through deep RL leaks much less information than the myopic LPPM for the same distortion level, indicating the benefits of considering all the history when taking actions at each time instant. This difference is less for $\boldsymbol{Q}_x^1$, since the temporal correlations in the location history is much less than $\boldsymbol{Q}_x^2$. Finally, both proposed and myopic LPPMs performances are the same for $\boldsymbol{Q}_x^0$, since the user movement with uniform distribution does not have temporal memory, and therefore, taking the history into account does not help.


\section{Conclusions}
We have studied the privacy-utility trade-off in LPPMs using mutual information as a privacy measure. Having identified some properties of the optimal policy, we recast the problem as an MDP. Due to continuous state and action spaces, it is challenging to characterize or even numerically compute the optimal policy. We overcome this difficulty by employing advantage actor-critic deep RL as a computational tool. Utilizing DNNs, we numerically evaluated the privacy-utility trade-off curve of the proposed location release policy. We compared the results with a myopic LPPM, and observed the effect of considering temporal correlations on information leakage-distortion performance. According to the simulation results, we have seen that the proposed LPPM policy provides significant privacy advantage, especially when the user trajectory has higher temporal correlations. 


\vspace{-0.22cm}
\bibliographystyle{IEEEtran}
\bibliography{IEEEexample}
%

\end{document}